\documentclass{article}

\usepackage[pdftex]{graphicx}
\usepackage{INTERSPEECH2022}

\title{Frequency-Directional Attention Model for Multilingual Automatic Speech Recognition}
\name{Akihiro Dobashi$^1$, Chee Siang Leow$^1$, Hiromitsu Nishizaki$^1$}
\address{
    $^1$Integrated Graduate School of Medicine, Engineering, and Agricultural Sciences,\\ University of Yamanashi, Japan
  }
 \email{\{dobashiakhiro, cheesiang\_leow\}@alps-lab.org, hnishi@yamanashi.ac.jp}

\begin{document}

\baselineskip 10.30pt 
\textfloatsep 3.3mm
\floatsep 3.0mm

\maketitle

%
\begin{abstract}
  This paper proposes a model for transforming speech features using the frequency-directional attention model for End-to-End (E2E) automatic speech recognition. 
  The idea is based on the hypothesis that in the phoneme system of each language, the characteristics of the frequency bands of speech when uttering them are different.
  By transforming the input Mel filter bank features with an attention model that characterizes the frequency direction, a feature transformation suitable for ASR in each language can be expected.
  This paper introduces a Transformer-encoder as a frequency-directional attention model.
  We evaluated the proposed method on a multilingual E2E ASR system for six different languages and found that the proposed method could achieve, on average, 5.3 points higher accuracy than the ASR model for each language by introducing the frequency-directional attention mechanism.
  Furthermore, visualization of the attention weights based on the proposed method suggested that it is possible to transform acoustic features considering the frequency characteristics of each language.
\end{abstract}

\section{Introduction}

In recent years, the introduction of deep learning has dramatically improved the recognition accuracy of automatic speech recognition systems (ASR). However, most research on ASR often focuses only on a single language speech. Therefore, for ASR systems that support multiple languages, it has been proposed to use a language-specific ASR system after a language identifier has been performed \cite{Kim/2020}. However, since this approach requires sophisticated language identifiers and language identification becomes difficult when short utterances or certain phrases or words are frequently code-switched to be replaced by another language, multilingual ASR methods that do not use language identifiers have also been studied. 
For example, a multilingual ASR system that performs language identification at the beginning of an utterance is being studied as a multilingual speech recognition system that does not use a language identifier \cite{Seki/2018}.
There is also an increasing amount of research dealing with code-switching, in which an utterance contains words in more than one language \cite{Zeng/2019}. 
However, training these neural network-based ASR models requires large amounts of multilingual speech data, which may make it impossible to build highly accurate speech recognition systems when the amount of training data is small.


Therefore, we have studied deep learning-based ASR that can handle multiple languages simultaneously with a single speech recognition model \cite{Hara/2017, Hayakawa/2021}.
For example, in our previous study \cite{Hara/2017}, we proposed a Hidden Markov Model (HMM)-Deep Neural Network (DNN) model \cite{Kaldi} using the International Phonetic Alphabet (IPA) model as a common language phoneme. However, the hybrid ASR model had many language identification errors and was often misrecognized as phonemes of other languages. In recent years, ASR has been shifting from a hybrid HMM-DNN approach to an end-to-end (E2E) approach, and the E2E framework is more favorable for multilingual simultaneous ASR models.
Hayakawa et al. \cite{Hayakawa/2021} proposed a method that attempts to extract language and speaker-universal acoustic features suitable for language and speaker-independent speech recognition in the E2E framework. This E2E model is a multitasking model that simultaneously performs language identification and speaker recognition in addition to speech recognition. By applying a gradient reversal layer (GRL) \cite{GRL} to the language identification and speaker recognition tasks, they have shown that the proposed method achieves language and speaker-independent acoustic feature extraction, improving simultaneous multilingual ASR accuracy.

\begin{figure*}[tb]
  \centering
  \includegraphics[width=2.0\columnwidth]{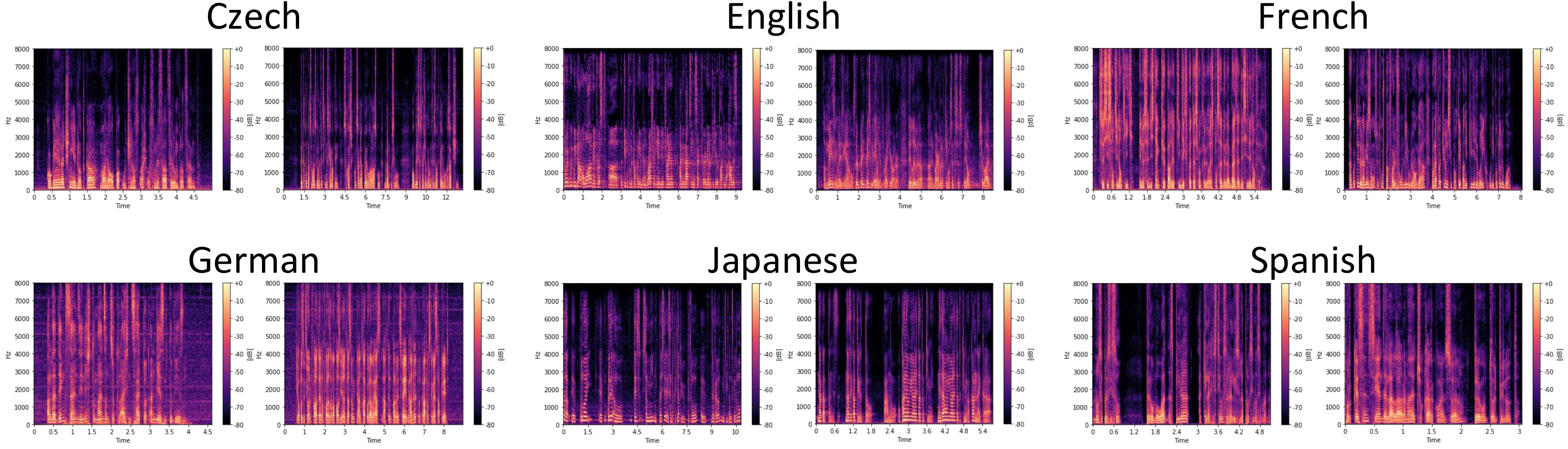}
  \caption{Speech Spectrograms for six languages}
  \label{fig:spectrogram}
\end{figure*}

Here, examples of spectrograms of spoken utterances in the six languages treated in the previous study \cite{Hayakawa/2021} are shown in Fig. \ref{fig:spectrogram}. These examples of spectrograms also show that different frequency bands are used depending on the phoneme system of each language. In the case of Japanese, the energy is concentrated in a relatively low-frequency band, while in French, the energy is distributed over a wide band. Assuming that the acoustic characteristics of the speech of each language are represented in the spectrogram, we can suppose that it is necessary to take into account such characteristics of the spectrogram in order to construct a multilingual ASR system with higher accuracy. In other words, if a framework that considers the characteristics of frequency bands can be introduced into a multilingual simultaneous ASR system, we believe that multilingual ASR using the characteristic information of each language can be realized.

\begin{figure*}[tb]
  \centering
  \includegraphics[width=1.8\columnwidth]{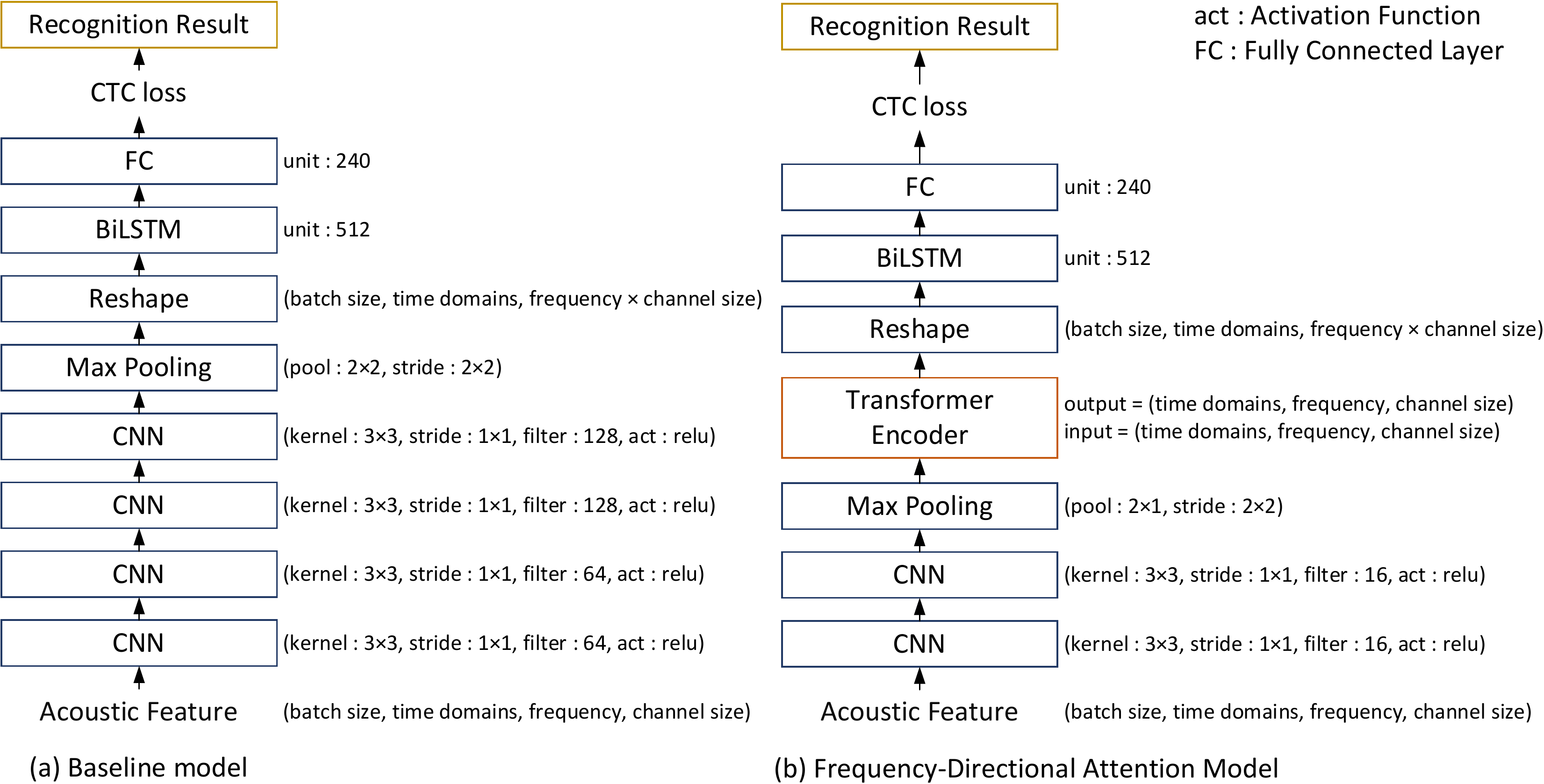}
  \caption{Multilingual E2E ASR models}
  \label{fig:model}
\end{figure*}

Therefore, this paper proposes a multilingual ASR model that uses a frequency-directional attention mechanism as a framework for successfully capturing frequency features in spectrograms. We show that the E2E model with the frequency-directional attention mechanism improves the accuracy of multilingual ASR. Although no study has been proposed to consider the frequency-directional attention mechanism in multilingual speech recognition, frequency-focused ASR models have been proposed for monolingual ASR. For example, Li et al. \cite{Li/2015}, proposed to input the features of the frequency bins at each time point into the  Long Short-Term Memory (LSTM) layer in the frequency direction before inputting the speech features into the LSTM layer in the time direction. Segbroeck et al.  \cite{Segbroeck/2020} modified this method to analyze speech under different speech analysis conditions and input the acoustic features extracted in each condition into the multi-stream frequency-axis LSTM layer. Although E2E ASR focusing on feature extraction in the frequency direction has been proposed, there are few examples of research on ASR models that apply an attention mechanism in the frequency direction. However, simultaneous temporal and frequency-directional attention mechanisms have been proposed in voice activity detection (VAD)\cite{Lee/2020} and speech enhancement processing \cite{Tang/2020}.

We propose the use of the Transformer \cite{transformer}, which has been introduced in E2E speech recognition in recent years with great success, as an attention mechanism in the frequency direction. A Transformer model has been introduced in ASR tools such as ESPnet \cite{espnet}, but most ASR models use them as an attention mechanism in the temporal direction, and no ASR system uses them as an attention mechanism in the frequency direction. In this study, each frame of the 40-dimensional logarithmic Mel filter bank (Mel-fbank) output is input to the Transformer-encoder, which performs a language-specific acoustic feature transformation and thus is expected to improve ASR accuracy.
Our E2E ASR model with the frequency-directional Transformer is trained and evaluated on six different languages in the multilingual ASR experiment. The experimental result showed that using the frequency-directional Transformer improved the average phoneme error rate of the six languages by 5.3 points (relatively 20\%) compared to the case without the frequency-directional Transformer. The result demonstrated the effectiveness of using the frequency-directional attention model in multiple ASR.

In consideration of the above, the contributions of this paper are as follows: 
\begin{itemize}
\item The first adaptation of a Transformer-based attention to the frequency direction in an E2E ASR model, and showing its effectiveness in improving the accuracy of multilingual ASR.
\end{itemize}

\begin{figure}[tb]
  \centering
  \includegraphics[width=0.95\columnwidth]{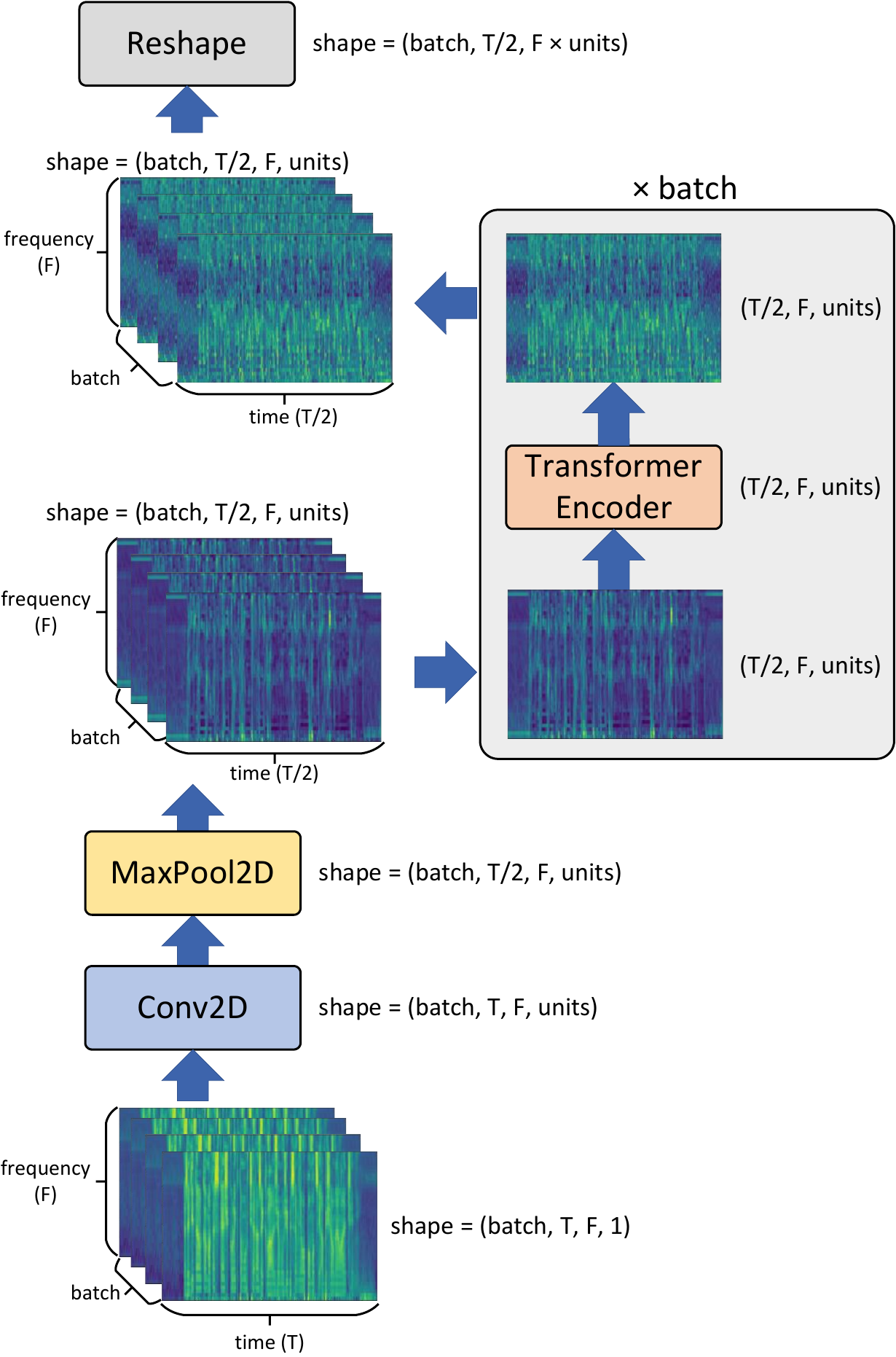}
  \caption{Architecture of the feature converter based on the frequency-directional attention model}
  \label{fig:transformer}
\end{figure}

\section{Frequency-Directional Attention Model} \label{sec:model}

In this study, we propose a multilingual ASR E2E model based on speech feature transformation focusing on frequency bands in order to build a single speech recognition system for multiple languages. Specifically, the E2E ASR model is based on a neural network that takes log Mel-fbank features as input, where the Connectionist Temporal Classification (CTC) \cite{ctc} is used as the loss function. A feature converter with an attention mechanism that characterizes each frequency bin of the log Mel-fbank features is connected to the ASR model.

Figure \ref{fig:model} shows the architecture of the baseline model and proposed model.
In the baseline model, the input features are transformed by four CNN layers. This is a very common method. After that, the CTC loss is computed through a Bidirectional LSTM (BiLSTM) layer and a fully-connected (FC) layer.
Figure \ref{fig:model}(b) shows the proposed model. Although the baseline model uses only CNN layers for feature conversion, the proposed model applies a Transformer-encoder-based frequency-directional attention mechanism to the output of the second CNN layers.

The Transformer-encoder introduced in this study has the following two characteristics.
\begin{itemize}
  \item Acoustic features at a certain time ($x_1, x_2, ..., x_n$, where $n$ is number of frequency bins) are transformed into an embedding vector ($z_1, z_2, ..., z_n$) by a multi-head self-attention mechanism.
  \item The self-attention mechanism refers to all frequency bins in the feature vector.
\end{itemize}
Using these characteristics, the Transformer-encoder directs attention to each frequency band of input acoustic features and perform an acoustic feature transformation.

The 40-dimensional log Mel-fbank outputs are used as acoustic features. Acoustic features based on the log Mel-fbank outputs have specific frequency band information in each dimension because of the way the features are extracted. Feature transformation is performed by directing attention to each frequency band with the Transformer-encoder. We assume that this will improve the language discrimination capability and improve the performance of the multilingual ASR model.

The proposed method needs to maintain temporal information of acoustic features in addition to frequency information, because the goal of this research is to improve the performance of the multilingual ASR task. However, using the Transformer-encoder in an E2E ASR model, an attention mechanism is usually directed only in the temporal direction.
Therefore, the input features are transformed, as shown in Fig. \ref{fig:transformer}. The mini-batch data is represented by the tensor shown in the lower left in Fig. 3. It is first input to the convolutional layer and then compressed in half in the time direction by the pooling layer behind the convolutional layer. Decomposing this with respect to the batch dimension, we can get the transformed feature represented by a 3-dimensional tensor of ($T/2, F, units$). In this study, $F$ and $units$ are 40 and 16, respectively.
By passing this through a Transformer-encoder layer, we can finally convert the feature with attention to the frequency direction. This process is repeated for the number of batch sizes and finally returned to the original tensor format, as shown in the upper left of Fig 3. This makes it possible to adapt the attention mechanism to the frequency band of all the data and still maintain temporal information.

\begin{table}[t]
  \centering
  \caption{The Dataset statistics and number of phoneme types in each language}
  \label{tbl:experiment_data}
  \vspace*{-2mm}
  \scalebox{1.0}{
  \begin{tabular}{c|c|c|c} \hline\hline
  language & training & test & number of phonemes  \\ \hline 
  Czech   & 17 [h] & 2.7 [h] & 41 \\
  English & 17 [h] & 2.6 [h] & 39 \\
  French  & 17 [h] & 2.0 [h] & 38 \\
  German & 17 [h] & 1.5 [h] & 41 \\
  Japanese &17 [h] & 5.1 [h] & 39 \\
  Spanish & 17 [h] & 1.7 [h] & 40 \\ \hline
  All & 102 [h] & 15.6 [h] & 238 \\ \hline\hline
  \end{tabular}
  }
\end{table}

\section{Experiments}
\subsection{Experimental Setup}

Table ~\ref{tbl:experiment_data} shows the statistics of datasets used in this paper.
We prepared six different languages speech corpora as follows:
\begin{itemize}
  \item Spanish, Czech, German and French from the GlobalPhone \cite{globalphone}
  \item English from the TED-LIUM corpus \cite{tedlium}
  \item Japanese from Corpus of Spontaneous Japanese \cite{csj}
\end{itemize}
In practice, the duration of training data differs for each speech corpus. In this study, the duration of training data for each language was standardized to 17 hours for German, which has the smallest amount of data. By unifying the duration of training data, we can confirm the effectiveness of the proposed method by removing the influence of the deterioration of ASR performance caused by the variation in the amount of training data.

The CTC-based E2E model has 240 output dimensions, including \verb|<blank>|, \verb|<unk>|, and 238 phonemes, as shown in Table \ref{tbl:experiment_data}. In other words, a phoneme recognition model is trained in this study. The reason for using the phoneme-based ASR model is to reduce the difference in the number of output symbol types for each language and evaluate whether the model can accurately capture differences in phonemes while identifying languages. For example, in Japanese, which includes Kanji characters, the number of output nodes is large (several thousand or more) due to a large number of character types, but by using phonemes, the conditions are almost the same for all languages. Therefore, the phoneme error rate (PER) is used as the evaluation measure. The output of the model is decoded into a sequence of phonemes using beam search and rescored using a phoneme-based 3-gram language model for all languages. The beam width was set to 20, and the probability weight of the language model was set to 1.0.

\begin{table}
  \centering
  \caption{Hyper-parameters for the model training}
  \label{tbl:parameter}
  \vspace*{-2mm}
  \begin{tabular}{c|c} \hline\hline
  Feature & Log Mel-fbank (40 dim.) \\
  Num. of epochs & 20 \\
  Mini-batchsize & 8 utterances \\ 
  Optimization & Adam (beta1:0.9, beta2:0.999) \\
  Dropout ratio & 0.1 \\
  Loss func. & CTC \\ \hline\hline
  \end{tabular}
  \vspace*{3mm}
  \end{table}
  
  \begin{table}[tb]
    \centering
    \caption{PERs [\%] in each model}
    \label{tbl:result}
    \vspace*{-2mm}
    \scalebox{1.0}{
    \begin{tabular}{c|c|c} \hline\hline
    language & baseline & proposed \\ \hline
    Czech & 20.9 & {\bf 17.2} \\
    English & 40.1 & {\bf 31.5} \\
    French & 22.5 & {\bf 15.7} \\
    German & 29.7 & {\bf 22.9} \\
    Japanese & 25.9 & {\bf 22.5} \\
    Spanish & 21.1 & {\bf 15.6} \\ \hline
    All languages & 26.6 & {\bf 21.3} \\ \hline\hline
    \end{tabular}
    }
    \end{table}
    
  \begin{figure*}[tb]
    \centering
    \includegraphics[width=1.95\columnwidth]{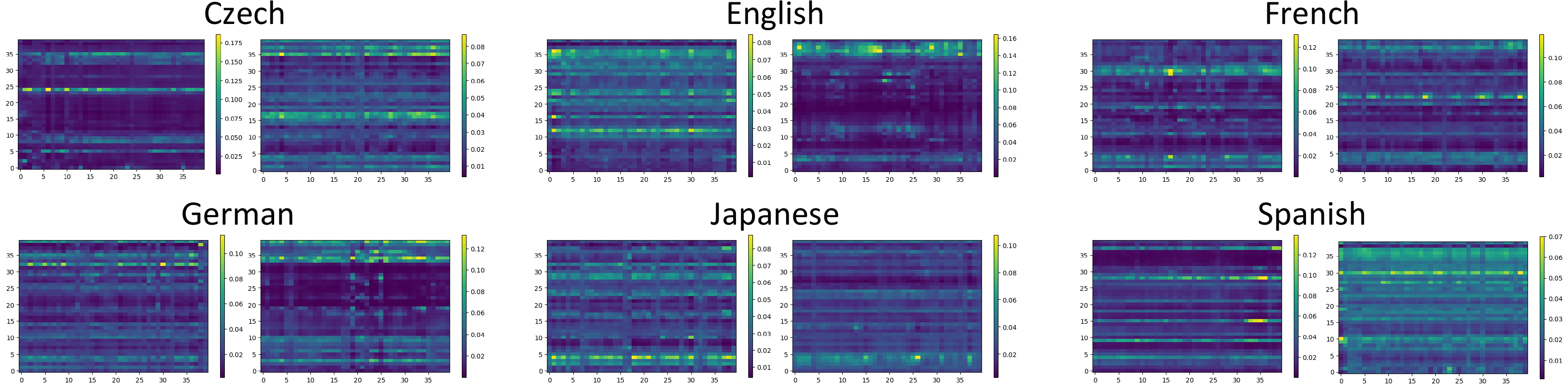}
    \caption{Visualization of frequency-directional attention weights for six languages}
    \label{fig:attention}
  \end{figure*}

\subsection{Training models}

To confirm whether the frequency-directional attention mechanism is an effective method in multilingual ASR models, we conduct a comparison experiment with a model that does not use the frequency-directional attention mechanism.

The models to be compared are as described in Section \ref{sec:model}, the Transformer Encoder has four layers and four multi-heads, the size of the d model is 16, and the dimension of the feed-forward network is 64. The parameter size of each model is smaller, about 4M for the proposed model compared to about 13M for the baseline model.

The hyper-parameters used for training are shown in Table \ref{tbl:parameter}. The learning rate was set to 0.0001 for the baseline model, and the proposed method used warm-up scheduling according to the following equation. Since the warm-up step was set to 5000, the learning rate follows equation (\ref{learning_rate_1}) when the warm-up step is less than 5000 steps and changes according to equation (\ref{learning_rate_2}) when the warm-up step is greater than 5000 steps.
\begin{equation}
  \label{learning_rate_1}
  lrate = 256^{-0.5} \times step\_num \times 5000^{-1.5}
\end{equation}
\vspace*{-4mm}
\begin{equation}
  \label{learning_rate_2}
  lrate = 256^{-0.5} \times step\_num^{-0.5}
\end{equation}

\subsection{Results}

Table ~\ref{tbl:result} shows the PERs of the evaluation set. The proposed method outperforms the baseline model without the frequency-directional attention mechanism by 5.3 points (relatively 20\%) in PER for all languages. The language-specific PERs also show improvement compared to the baseline. This shows that adding the Transformer attention mechanism in the frequency direction improves the performance in multilingual ASR.

Figure \ref{fig:attention} is a heat map visualizing the weights of the attention mechanisms in a certain section of speech data for each language (two sections for each language). The vertical axis shows the input features (40-dimensional log Mel-fbank output), and the horizontal axis shows the Transformer-encoder output (40 dimensions). For example, we can see that Japanese, which uses many vowels, pays more attention to the low-frequency band, while the Latin languages, which use many consonants, pay more attention to the high-frequency band. Thus, the Transformer-encoder for the frequency direction is used to perform feature conversion that captures the characteristics of the language. This is considered to contribute to the improvement of multilingual ASR performance.



\section{Conclusions}

This paper proposed a novel method to introduce a frequency-directional attention mechanism into the CTC loss-based E2E model for multilingual ASR. Transformer-encoder is used as the attention mechanism, which converts log Mel-fbank features, and it achieved an ASR model that takes advantage of the characteristics of each language. The results of multilingual ASR experiments showed that the proposed model significantly outperformed the ASR accuracy of the baseline model. In addition, by visualizing the weights of the attention mechanism, we confirmed that the feature transformation was performed based on the characteristics of each language. These results indicate that the frequency-directional attention mechanism is useful in multilingual ASR.

In the future, we are going to evaluate an improved ASR model that simultaneously applies an attention mechanism for both the temporal direction and the frequency direction to the attention mechanism. In addition, it is possible that this method can also make good use of the characteristics of each speaker's speech, instead of using multilingual speech recognition. For example, we would like to evaluate this method for the elderly speech that is difficult to recognize. 

\section{Acknowledgements}
This work was supported by JSPS KAKENHI Grant Number 21H00901. 
Besides, a part of this work was also supported by the Hoso Bunka Foundation.

\newpage
\bibliographystyle{IEEEtran}

\bibliography{main}


\end{document}